\documentclass[aps,prb,reprint,english,a4paper,10pt,floatfix,superscriptaddress,showpacs]{revtex4-1}
\usepackage[T1]{fontenc}
\usepackage{ae,aecompl}
\usepackage{courier}
\usepackage[english]{babel}
\usepackage[normalem]{ulem}
\usepackage{url}
\usepackage{psfrag,color,pstricks,pst-grad}
\usepackage[dvips]{graphicx}
\usepackage{amsmath,amsfonts,amssymb,amsbsy}  
\usepackage{hyperref}

\newcommand{\bb}{\mathbf}

\newcommand{\ca}{\mathcal}

\newcommand{\begeq}{\begin{equation}}

\begin{document}
 
  \title{Frequency-dependent phonon mean free path in carbon nanotubes from non-equilibrium molecular dynamics}
\author{K. S\"a\"askilahti}
\email{kimmo.saaskilahti@aalto.fi}
\affiliation{Department of Biomedical Engineering and Computational Science, Aalto University, FI-00076 Aalto, Finland}

\author{J. Oksanen}
\affiliation{Department of Biomedical Engineering and Computational Science, Aalto University, FI-00076 Aalto, Finland}

\author{S. Volz}
\email{sebastian.volz@ecp.fr}
\affiliation{Ecole Centrale Paris, Grande Voie des Vignes, 92295 Ch\^atenay-Malabry, France}
\affiliation{CNRS, UPR 288 Laboratoire d'Energ\'etique Mol\'eculaire et Macroscopique, Combustion (EM2C), Grande Voie des Vignes, 92295 Ch\^atenay-Malabry, France}

\author{J. Tulkki}
\affiliation{Department of Biomedical Engineering and Computational Science, Aalto University, FI-00076 Aalto, Finland}

\date{\today}

\begin{abstract}
Owing to their long phonon mean free paths (MFPs) and high thermal conductivity, carbon nanotubes (CNTs) are ideal candidates for, e.g., removing heat from electronic devices. It is unknown, however, how the intrinsic phonon MFPs depend on vibrational frequency in \textit{non-equilibrium}. We determine the spectrally resolved phonon MFPs in isotopically pure CNTs from the spectral phonon transmission function calculated using non-equilibrium molecular dynamics, fully accounting for the resistive phonon-phonon scattering processes through the anharmonic terms of the interatomic potential energy function. Our results show that the effective room temperature MFPs of low-frequency phonons ($f<0.5$ THz) exceeds $10$ $\mu$m, while the MFP of high-frequency phonons ($f\gtrsim 20$ THz) is in the range 10--100 nm. Because the determined MFPs directly reflect the resistance to energy flow, they can be used to accurately predict the thermal conductivity for arbitrary tube lengths by calculating a single frequency integral. The presented results and methods are expected to significantly improve the understanding of non-equilibrium thermal transport in low-dimensional nanostructures. 
 \end{abstract}
 \maketitle

\section{Introduction}

The small atomic mass, rigid $sp^2$-bonding and high structural order of carbon atoms in carbon nanotubes and graphene gives rise to exceptional mechanical and thermal properties. For example, the thermal conductivity (TC) of carbon nanotubes (CNTs) has been theoretically predicted \cite{berber00,che00,osman01,mingo05,donadio07,cao12,salaway14} and experimentally measured \cite{kim01,yu05,pop05,pop06} to be in the range 500--7000 W/(mK) at room temperature, depending on, e.g., the tube chirality, tube length and experimental setups \cite{marconnet13}. The high TC, mechanical strength and the vast possibilities of chemical functionalization make carbon nanotubes attractive for applications aiming at, e.g., efficient heat removal in electronics \cite{kaur14}, thermoelectric interface materials \cite{prasher09,gao10} and phonon waveguides and rectifiers \cite{chang06_prl,chang06}. 

Heat is carried in CNTs primarily by the propagating lattice vibrations, phonons \cite{balandin11}. One of the key unknown factors determining the phononic TC in isotopically pure nanotubes is the mean free path (MFP) describing the characteristic distance of resistive phonon-phonon scattering events. The effective MFP in CNTs has been experimentally estimated to be in the range 500--750 nm at room temperature \cite{kim01,yu05}. These estimates are based, however, on a kinetic formula, which is known to underestimate the true MFP \cite{ju99} and cannot account for the strong dependence of MFP on the phonon frequency \cite{klemens94,cao04,mingo05_nanolett,gu07,lindsay09}. The frequency-dependence is visible, e.g., in the thermal conductivity accumulation function \cite{dames06, dames13}, whose experimental measurement has been recently enabled by advanced spectroscopic techniques \cite{minnich11a,regner13,johnson13,cuffe14}.

The frequency-dependent MFPs have been previously determined theoretically either from the decay of the mode energy correlation function \cite{ladd86,mcgaughey04} in equilibrium molecular dynamics (EMD) simulations \cite{donadio07,ong11}  or by calculating the phonon-phonon scattering rates from first principles \cite{xiao03,cao04,mingo05_nanolett,hepplestone06,gu07,lindsay09}. While it is known that only the Umklapp scattering processes can directly generate thermal resistance due to the change in crystal momentum \cite{ziman,Note1}, both normal and Umklapp scattering processes contribute to redistributing the mode energy. Therefore, the MFPs obtained from EMD simulations do not directly reflect the decay of the heat flux in non-equilibrium \cite{ziman}. In first principles calculations, on the other hand, typically only three-phonon scattering in the first order is considered. Higher-order scattering processes and four-phonon scattering processes are either neglected or treated approximately \cite{mingo05_nanolett}. In addition, determining TC from the first-principles scattering rates requires \cite{lindsay09} the solution of the highly complicated Boltzmann equation.

In this article, we provide an in-depth evaluation of the frequency-dependence of the effective phonon MFPs and phonon transmission functions based on non-equilibrium molecular dynamics (NEMD) simulations and the generalized form of the recently developed expression for the spectral decomposition of the non-equilibrium heat current \cite{saaskilahti14b}. The obtained phonon MFPs reflect the length-dependence of the transmission function arising from the phonon-phonon interactions implicitly included in our simulations through the anharmonic terms in the interatomic potential energy function.  Because the non-equilibrium heat current inherently accounts for the different roles of normal and Umklapp processes in generating thermal resistance, the determined MFPs capture the subtle interplay of normal and Umklapp processes. In contrast to first-principles calculations of MFPs, NEMD also accounts for all orders of all phonon-phonon interactions and the effective MFPs can be directly used to predict TC, as shown below.

The paper is organized as follows. We first generalize the recently developed expression for the spectral heat current to describe many-body potentials in Sec. \ref{sec:spectralformula}. We then define the generalized phonon transmission function and show how the MFPs can be determined from the length-dependence of the transmission in Sec. \ref{sec:transmission}. In Sec. \ref{sec:ktheta}, we discuss the decomposition of the transmission function into different angular wavenumber contributions, allowing for calculating the relaxation times for different wavenumbers and comparison to other methods. The molecular dynamics setup and the numerical methods are presented in Sec. \ref{sec:setup}, followed by a detailed overview of the numerically calculated length-dependent transmission functions and mean free paths in Sec. \ref{sec:results} and their dependence on the angular wavenumbers and comparison to EMD in Sec. \ref{sec:ktheta_results}.

\section{Theory} 
\label{sec:theory}
\subsection{Spectral heat current formula}
\label{sec:spectralformula}
To calculate the phonon MFPs from the decrease of the phonon transmission function as a function of tube length, we need to evaluate the phonon transmission function from the anharmonic NEMD simulations. To achieve this, we first need to generalize our previous approach \cite{saaskilahti14b} to calculate the spectrally resolved heat current to many-body potentials such as the Tersoff potential. Here we briefly overview this generalization by complementing the previous derivation of Ref. \cite{saaskilahti14b}.

The expression for the interparticle heat current $Q_{i\to j}$ between atoms $i$ and $j$ (located at different sides of the imaginary cross-section) is given by \cite{lepri03,narayan09}
\begin{equation}
 Q_{i\to j} = \frac{1}{2} \langle \mathbf{F}_{ji}\cdot \mathbf{v}_j-\mathbf{F}_{ij}\cdot \mathbf{v}_i\rangle. \label{eq:Qji}
\end{equation}
Here, the angular brackets denote the steady-state non-equilibrium ensemble average assumed to be equal to the time average due to ergodicity, the velocities of atoms $i$ and $j$ are denoted by $\mathbf{v}_i$ and $\mathbf{v}_j$, respectively, and $\mathbf{F}_{ji}$ is the interparticle force on atom $j$ due to atom $i$. The spectral heat current $q_{i\to j}(\omega)$, satisfying $Q_{i\to j} =\int_0^{\infty}(d\omega/2\pi) q_{i\to j}(\omega)$ with $\omega$ the angular frequency, was shown to be given by the expression \cite{saaskilahti14b}
\begin{equation}
  q_{i\to j} (\omega) = 2 \textrm{Re} [\tilde K_{ji}(\omega)], \label{eq:qij}
\end{equation}
where $\tilde{K}_{ji}(\omega)=\int_{-\infty}^{\infty} dt e^{i\omega t} K_{ji}(t) $ is the Fourier transform of the force-velocity cross-correlation function defined in time-domain as
\begin{equation}
 K_{ji}(t_1-t_2) = \frac{1}{2} \langle \mathbf{F}_{ji}(t_1) \cdot \mathbf{v}_j(t_2)-\mathbf{F}_{ij}(t_1)\cdot \mathbf{v}_i(t_2)\rangle. \label{eq:Kji}
\end{equation}
The correlation function \eqref{eq:Kji} depends explicitly only on the time difference $t_1-t_2$ (and the Fourier transform on a single frequency variable) due to the assumed steady state.

In contrast to Ref. \cite{saaskilahti14b}, where the forces were generated due to two-particle interactions (more specifically Lennard-Jones potentials), in the present CNT systems the forces have also three-body contributions. Therefore, it is necessary to carefully define the interatomic forces $\bb{F}_{ji}$ used in analyzing the force-velocity correlation function \eqref{eq:Kji} using NEMD.

The required expression for the interparticle force can be derived from the interparticle heat current, which can in turn be calculated by monitoring the temporal rate of change of the local energy \cite{hardy63,guajardo10}. The system's total energy $E=K+U$, which consists of the kinetic energy $K$ and potential energy $U$, is written as the sum over the local energies $\varepsilon_i$ of atoms $i\in\{1,\dots,N\}$:
\begin{equation}
 E= \sum_i \varepsilon_i.
\end{equation}
The local energy $\varepsilon_i$ consists of the local kinetic and potential energy contributions,
\begin{equation}
\varepsilon_i =  \frac{1}{2} m \mathbf{v}_i^2 + U_i(\mathbf{r}_1,\dots,\mathbf{r}_N) ,
\end{equation}
where $\mathbf{r}_i$ is the atomic position, $m$ is the atom mass, $N$ is the number of atoms and $U_i$ the local potential energy. By using the equation of motion
\begin{equation}
 m\frac{d\mathbf{v}_i}{dt} = - \frac{\partial U}{\partial \mathbf{r}_i},
\end{equation}
one can show that the temporal rate of change of local energy, which equals the in-flow of heat current, is
\begin{alignat}{2}
 \frac{d\varepsilon_i}{dt} & = - \sum_{j\neq i} \left(- \frac{\partial U_i}{\partial \mathbf{r}_j} \cdot {\mathbf{v}}_j +  \frac{ \partial U_j }{\partial \mathbf{r}_i} \cdot {\mathbf{v}}_i \right). \label{eq:dedt}
\end{alignat}
Identifying the term in parentheses on the right-hand side as the interparticle heat current and comparing to Eq. \eqref{eq:Qji} leads to conclude that for general interatomic potentials, the interparticle force $\bb{F}_{ji}$ used to calculate the heat currents is given by
\begin{equation}
 \mathbf{F}_{ji} = - 2\frac{\partial U_i}{\partial \mathbf{r}_j}. \label{eq:Fji}
\end{equation}
 
The spectral heat current \eqref{eq:qij} can be calculated using force and velocity trajectories from NEMD simulation with fully anharmonic interatomic potentials. Keeping track of the generalized interparticle forces \eqref{eq:Fji} is, however, complicated and tedious. Therefore, it is useful to derive an expression for the spectral heat current that only requires the atomic velocity trajectories instead of explicit interparticle forces. For a solid, this can be achieved by expanding the interparticle force in terms of small atomic displacements $\mathbf{u}_i=\mathbf{r}_i-\mathbf{r}_i^0$ from the average position $\mathbf{r}_i^0$. It is important to emphasize that this expansion is only applied in the post-processing phase to simplify the calculation of the spectral heat current. Fully anharmonic forces, accounting for all orders of phonon-phonon interactions, are used in the NEMD simulation.

The first-order term of Eq. \eqref{eq:qij}, which turns out to be the strongly dominant term for the rigid carbon-carbon interactions considered in this article, is obtained by approximating the total potential energy as the quadratic sum
\begin{equation}
 U \approx \frac{1}{2} \sum_{i,j} \sum_{\alpha,\beta}  {u}_i^{\alpha} K_{ij}^{\alpha\beta} u_j^{\beta}, \label{eq:U_quadratic}
\end{equation}
where the force constant matrix is
\begin{equation}
 K_{ij}^{\alpha\beta} = \left. \frac{\partial^2 U}{\partial u_i^{\alpha} \partial u_{j}^{\beta}} \right|_{\mathbf{u}=\mathbf{0}}. \label{eq:Kij_def}
\end{equation}
Here, the co-ordinates are $\alpha,\beta\in\{x,y,z\}$. From Eq. \eqref{eq:U_quadratic}, one can see that the local potential energy $U_i$ can be approximated by
\begin{equation}
 U_i \approx \frac{1}{2} \sum_j \sum_{\alpha,\beta}  {u}_i^{\alpha} {K}_{ij}^{\alpha\beta} {u}_j^{\beta}.
\end{equation}
This results in the generalized interparticle force [Eq. \eqref{eq:Fji}, $j\neq i$]
\begin{equation}
 F_{ji}^{\gamma} \approx - \sum_{\alpha} u_i^{\alpha} K_{ij}^{\alpha\gamma}.
\end{equation}

One can then use Fourier transform identities and the correspondence between continuous and discrete Fourier transforms to show that the spectral heat current \eqref{eq:qij} can be calculated from the compact expression
\begin{equation}
 q_{i \to j}(\omega) \approx -\frac{2}{t_{\textrm{simu}} \omega} \sum_{\alpha,\beta\in\{x,y,z\}} \textrm{Im}\left.\left\langle \hat{v}_i^{\alpha}(\omega)^* K_{ij}^{\alpha\beta} \hat{v}_j^{\beta}(\omega) \right\rangle \right. \label{eq:qomega_expr}.
\end{equation}
Here $t_{\textrm{simu}}$ is the simulation time and the velocities $\hat v_i^{\alpha}(\omega)$ and $\hat v_j^{\beta}(\omega)$ are the discrete Fourier transforms of atomic velocity trajectories $v_i^{\alpha}(t)=\dot{u}_i^{\alpha}(t)$ (more details below in Sec. \ref{sec:setup}). The heat current across any interface separating disjoint atom sets $\tilde L$ and $\tilde R$, which we will choose to be the left and right halves of the tube, is obtained by summing over atoms in each set:
\begin{equation}
 q(\omega) = \sum_{i\in \tilde L} \sum_{j\in \tilde R} q_{i \to j}(\omega). \label{eq:qomega_sum}
\end{equation}

\subsection{Transmission and mean free paths}
\label{sec:transmission}

Having expressions \eqref{eq:qomega_expr}--\eqref{eq:qomega_sum} for the spectral heat current available, we define the generalized phonon transmission function as
\begin{equation}
 \ca{T}(\omega) = \frac{q(\omega)}{k_B \Delta T} \label{eq:Tomega_def}.
\end{equation}
In the ballistic limit, this transmission function is simply equal to $M(\omega)$, the number of propagating modes \cite{rego98}, which can be determined from the phonon bandstructure. The anharmonic phonon-phonon interactions incorporated in NEMD render the transmission function \eqref{eq:Tomega_def} dependent on the tube length $L$, which can be phenomenologically taken into account through the relation \cite{datta,wang06_apl,yamamoto09,saaskilahti13}
\begin{equation}
 \ca{T}(\omega) = \frac{M(\omega)}{1+L/\Lambda(\omega)}, \label{eq:T_diffusive}
\end{equation}
where $\Lambda(\omega)$ is the effective phonon MFP. Equation \eqref{eq:T_diffusive} smoothly interpolates between the ballistic [$\ca{T}(\omega)=M(\omega)$] and diffusive [$\ca{T}(\omega) \sim1/L$] limits. It can be derived by treating the phonon-phonon scattering events as resistance sources and combining the resistances incoherently \cite{datta} or from the Boltzmann transport equation under the frequency-dependent relaxation time approximation \cite{saaskilahti13}. Note that the mean free paths of Ref. \cite{saaskilahti13} correspond to the decay length of the phonon density, proportional to the square $|u|^2$ of the phonon amplitude, whereas our MFP definition corresponds to the decay in phonon amplitude $u$. Hence the mean free paths differ by a factor of two. Both definitions are correct as long as one remains consistent when using the MFPs to calculate thermal properties.

Equation \eqref{eq:T_diffusive} has been previously used \cite{wang06_apl,yamamoto09} to develop simplified models for the ballistic-diffusive transition in CNTs. Here, in contrast, we use Eq. \eqref{eq:T_diffusive} to determine the mean free paths from the relation between $\ca{T}(\omega)$, $M(\omega)$ and tube length $L$. This procedure was also used by Savic, Mingo and Stewart \cite{savic08_prl} to determine the impurity scattering MFPs in CNTs, but the harmonic Green's function simulations they used could not account for the phonon-phonon interactions incorporated in our simulations through the anharmonic terms in the interatomic potential. 

From the spectral heat current $q(\omega)$, we also calculate the spectral decomposition of the thermal conductivity $\kappa=Q/(A|dT/dx|)$ as
\begin{equation}
 \kappa(\omega) = \frac{q(\omega)}{A\Delta T} L. \label{eq:kappa_decomp_q}
\end{equation}
In the definition of $\kappa$, $Q$ is the total heat current flowing along the tube and $A$ is the cross-sectional area, typically defined \cite{marconnet13} for the hollow tube as $A=\pi d \times 0.34$ nm, where $d$ the tube diameter. In Eq. \eqref{eq:kappa_decomp_q}, we assumed $dT/dx\approx -\Delta T/L$ for the temperature gradient to allow for predicting the thermal conductivity for tubes of arbitrary length by utilizing Eqs. \eqref{eq:Tomega_def} and \eqref{eq:T_diffusive} and the MFPs:
\begin{equation}
 \kappa(\omega)= \frac{k_BL}{A} \frac{M(\omega)}{1+L/\Lambda(\omega)} . \label{eq:kappa_decomp}
\end{equation}

\subsection{Wavenumber decomposition of the transmission function}
\label{sec:ktheta}
At each frequency $\omega$, there are typically multiple propagating phonon modes with different polarizations and wavevectors. These degenerate modes may have different mean free paths due to the different probabilities for the multiple scattering events. Therefore, the MFPs determined from Eq. \eqref{eq:T_diffusive} correspond to an \textit{average} or effective scattering length at each frequency. To separately derive the mean free paths of different phonon branches and to simplify the determination of the relaxation times from the mean free paths, we decompose the transmission function \eqref{eq:Tomega_def} into different angular wavenumber contributions.

Owing to the rotational symmetry of the nanotube, the phonon states in a $(n,n)$ nanotube can be labeled by their angular wavenumber $k_{\theta} \in \{-n/2+1,\dots,n/2\}$ \cite{popov00,mahan04} (we assume throughout that $n$ is even). The label $k_{\theta}$ signifies the dependence of the phonon amplitude $u\sim \exp(ik_{\theta} \theta)$ on the azimuthal angle $\theta$ along the tube circumference. The decomposition of the transmission function into $k_{\theta}$-components relies on decomposing the spectral heat current into its angular wavenumber components as $q(\omega)=\sum_{k_{\theta}} \tilde q^{}(\omega,k_{\theta})$. This procedure, outlined in App. \ref{app:ktheta}, is similar to the one presented in Ref. \cite{chalopin13}, where the authors decomposed the transmission function in terms of the in-plane wavevectors at solid-solid interfaces. There are, however, some differences arising from the fact that that the tubes exhibit rotational invariance instead of translational invariance.

From the decomposed spectral current, the wave-vector decomposed transmission function is obtained as
\begin{equation}
 \ca{T}(\omega,k_{\theta}) = \frac{\tilde q^{}(\omega,k_{\theta})}{k_B \Delta T}. \label{eq:T_ktheta}
\end{equation}
The wavenumber decomposed mean free paths $\Lambda(\omega,k_{\theta})$ can then be determined from the formula analogous to Eq. \eqref{eq:T_diffusive}:
\begin{equation}
 \ca{T}(\omega,k_{\theta}) = \frac{M(\omega,k_{\theta})}{1+L/\Lambda(\omega,k_{\theta})}. \label{eq:mfp_ktheta}
\end{equation}
Here $M(\omega,k_{\theta})$ is the number of modes for given $\omega$ and $k_{\theta}$, determined from the wavenumber-decomposed bandstructure shown below in Sec. \ref{sec:ktheta_results} for a (10,10) nanotube.

\section{Numerical results} 

\subsection{Molecular dynamics setup}
\label{sec:setup}
\begin{figure}[tb]
 \begin{center}
  \includegraphics[width=8.6cm]{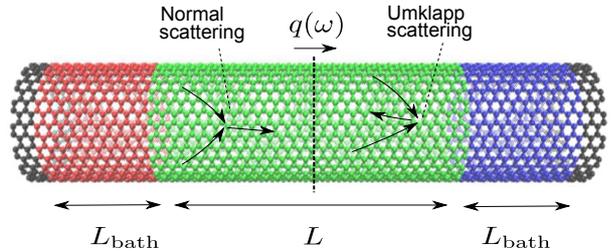}  
  \caption{(Color online) Schematic illustration of the NEMD setup and phonon-phonon scattering processes. Thermal current in a CNT is generated by coupling regions of length $L_{\textrm{bath}}$ to Langevin heat baths at temperatures $T+\Delta T/2$ and $T-\Delta T/2$ at the left and right ends of the tube, respectively. Inside the unthermalized region of length $L$, phonons traveling between the heat baths undergo phonon-phonon scattering processes. Whereas the normal scattering processes only redistribute phonon energies, the Umklapp processes generate thermal resistance, giving rise to the length-dependence of the spectral heat current $q(\omega)$ evaluated at the middle cross-section of the tube (thick dashed line).}  
\label{fig:setup}
 \end{center}
\end{figure}

For the numerical results, we study thermal conduction in a single-walled CNT using the computational NEMD setup schematically depicted in Fig. \ref{fig:setup}. To generate the heat current through the tube, the atoms located within the distance $L_{\textrm{bath}}$ from the left and right ends of the tube are coupled to hot and cold Langevin heat baths at temperatures $T+\Delta T/2$ and $T-\Delta T/2$, respectively. Two left and right-most translational unit cells are maintained at fixed positions to prevent large deformations and moving of the tube. The setup eliminates the contact resistance between the tube and the heat baths when the bath coupling time constant $\tau_{\textrm{bath}}$ is small and $L_{\textrm{bath}}$ is large \cite{saaskilahti14b}, ensuring that the spectral heat current is only limited by the internal conductance of the nanotube and does not depend on $\tau_{\textrm{bath}}$ and $L_{\textrm{bath}}$. 

The numerical results will be calculated for an armchair CNT with (10,10) chirality. The diameter of the (10,10) tube is $d=1.36$ nm. The optimized Tersoff potential \cite{lindsay10} is used for modeling the interatomic interactions. The simulation time step is $0.5$ fs, the duration of the data collection run $t_{\textrm{simu}}=25$ ns, the bath coupling time constant is chosen as $\tau_{\textrm{bath}}=1$ ps and the length of the thermalized regions $L_{\textrm{bath}}=36$ nm. The mean bath temperature is fixed at $T=300$ K and the values of temperature bias $\Delta T$ are $\Delta T=60$ K, $\Delta T=100$ K, and $\Delta T=200$ K for $L \leq 800$ nm, $L \in \{1,2\}$ $\mu$m, and $L=4$ $\mu$m, respectively. We have checked that the used values for $\Delta T$ are small enough to keep heat transfer in the linear regime by confirming that the transmission functions remain practically unchanged when the bias in reduced. The simulations are performed using the \texttt{LAMMPS} simulation package \cite{plimpton95,lammps_website}. 

We calculate the spectral heat current \eqref{eq:qomega_expr}--\eqref{eq:qomega_sum} across the cross-section located at the middle of the tube, depicted by the thick dashed line in Fig. \ref{fig:setup}. We have checked that the spectral current is insensitive to the exact position of the cross-section by calculating the current spectra flowing across cross-sections located at different positions along the tube and noting that the spectra agree. For the calculation of $q(\omega)$, the interatomic force constants $K_{ij}^{\alpha\beta}$ appearing in Eq. \eqref{eq:qomega_expr} are determined from the finite-difference derivatives of the interatomic potential energy function. In the simulation run, the velocities $v_i^{\alpha}(t)=\dot{u}_i^{\alpha}(t)$ of atoms located within the potential cut-off distance from the cross-section in the middle of the tube are sampled at intervals $\Delta t_s=5$ fs for the duration of the simulation $t_{\textrm{simu}}=N_f\Delta t_s$. The trajectories are used to evaluate the discrete Fourier transforms
\begin{equation}
 \hat{v}_i^{\alpha}(\omega_m) = \Delta t_s \sum_{k=0}^{N_f-1} e^{i\omega_m k \Delta t_s} v_i^{\alpha}(k\Delta t_s) 
\end{equation}
at the discrete frequencies $\omega_m=2\pi m /(N_f \Delta t_s)$, $m=0,1,\dots,N_f-1$. The discrete Fourier transforms are then used to calculate the spectral heat current [Eqs. \eqref{eq:qomega_expr}--\eqref{eq:qomega_sum}] flowing across the middle cross-section.  The obtained sharply fluctuating spectral heat current is smoothened by convolving with a Gaussian window with standard deviation $\Delta f=0.1$ THz. We have checked that the anharmonic contribution to the spectral heat current, disregarded in Eq. \eqref{eq:qomega_expr}, is negligible by comparing the integral of Eq. \eqref{eq:qomega_expr} to the total heat current $Q$ (determined from the work done by the heat baths) and confirming that the values agree up to statistical accuracy.  

\subsection{Spectral transmission and mean free paths}
\label{sec:results}
\begin{figure}[tb]
 \begin{center}
  \includegraphics[width=8.6cm]{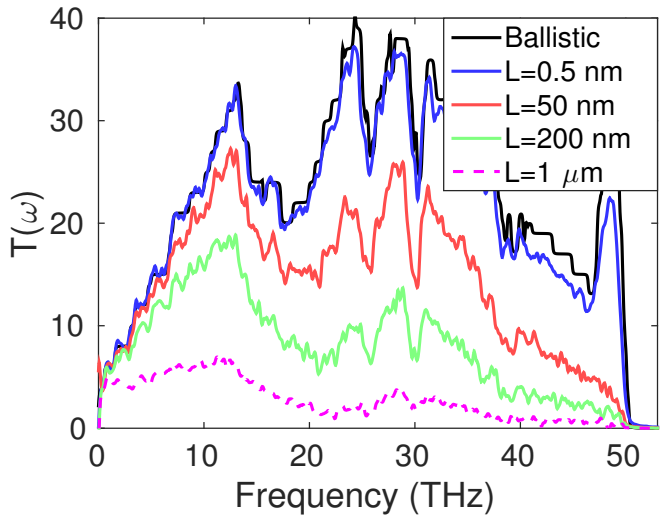} 
  \caption{(Color online) Spectral transmission function $\ca{T}(\omega)=q(\omega)/(k_B\Delta T)$ for various tube lengths at $T=300$ K, determined from the NEMD simulations. As expected, increasing the tube length reduces the transmission. For $L=0.5$ nm, the spectral conductance is very close to the ballistic value $M(\omega)$ determined by counting the number of propagating modes from the phonon bandstructure of Fig. \ref{fig:bs}.}  
\label{fig:Tom}
 \end{center}
\end{figure}

Figure \ref{fig:Tom} shows the transmission function \eqref{eq:Tomega_def} for four different tube lengths $L=0.5$ nm, $L=50$ nm, $L=200$ nm, and $L=1$ $\mu$m. In the shortest tube, the unthermalized part of the tube consists of only two translational unit cells, corresponding to the length $L \approx 0.5$ nm. The transmission through such a short tube is, as expected, practically equal to the ballistic value $M(\omega)$, the number of propagating modes in a (10,10) CNT (black, solid line), showing that the heat current flowing in the tube is only limited by the number of propagating modes and not by contact resistance to heat baths. 

As the tube length is increased to $L=50$ nm, the transmission function decreases significantly due to phonon-phonon scattering. The decrease in the transmission is especially strong in the high frequencies due to the large available phase-space for phonon-phonon scattering and the small group velocity. For $f\lesssim 5$ THz, the transmission is still nearly equal to the ballistic value for $L=50$ nm, suggesting that MFP in this frequency range is longer than 50 nm. For $L=1$ $\mu$m, the transmission is close to ballistic below $1$ THz, implying that such low-frequency modes can propagate ballistically even through a 1-$\mu$m-long tube.

\begin{figure}[tb]
 \begin{center}
  \includegraphics[width=8.6cm]{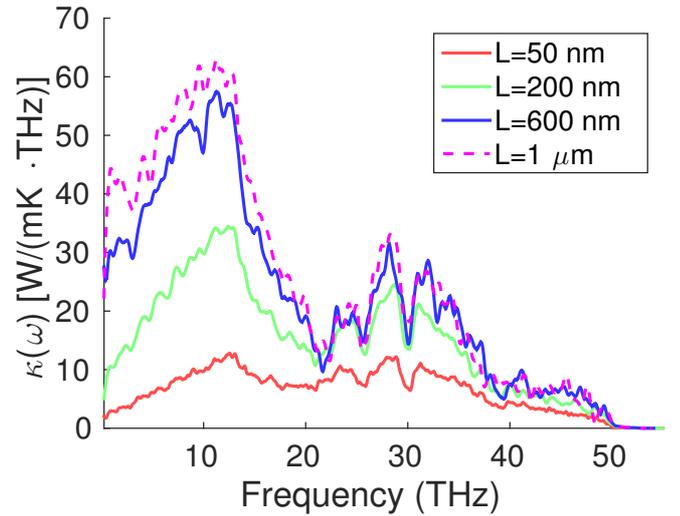} 
  \caption{(Color online) Spectral thermal conductivity $\kappa(\omega)=q(\omega)L/(A\Delta T)$ for various tube lengths at $T=300$ K, determined from the NEMD simulations. The absolute contribution of high frequencies ($f\gtrsim 15$ THz) has converged as a function of length for $L=1$ $\mu$m due to the short mean free paths. The absolute contribution of low-frequency vibrations, on the other hand, increases because of the (partially) ballistic transport.}  
\label{fig:kappaom}
 \end{center}
\end{figure}

Figure \ref{fig:kappaom} shows the spectral decomposition \eqref{eq:kappa_decomp_q} of thermal conductivity for various tube lengths. In the shortest tubes, the spectral conductivity increases as a function of tube length in the whole frequency range due to ballistic phonon transport. Once the tube length exceeds the mean free path, phonon transport becomes more diffusive and the spectral conductivity eventually converges. In Fig. \ref{fig:kappaom}, this convergence can be observed for the longest tubes for $f\gtrsim 12$ THz, suggesting that the MFPs in this frequency range are markedly below $600$ nm. At low frequencies, the conductivity still significantly increases, suggesting that the MFPs are in the micrometer range.

\begin{figure}[tb]
 \begin{center}
  \includegraphics[width=8.6cm]{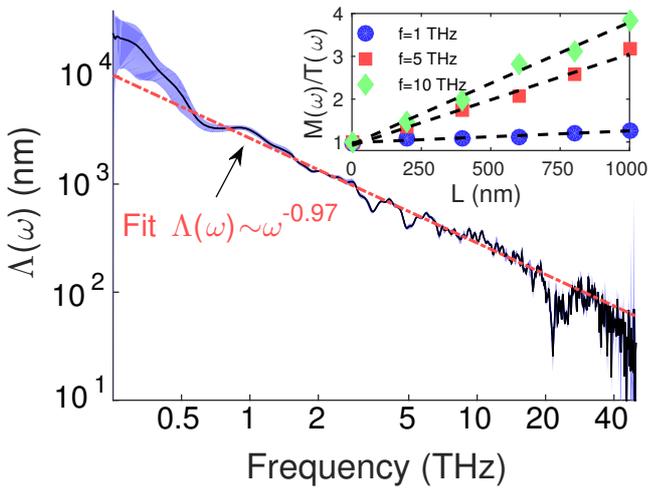}
  \caption{(Color online) Log-log plot of the mean free path $\Lambda(\omega)$ at $T=300$ K. The inset shows the scaled inverse transmission functions $M(\omega)/\ca{T}(\omega)$ as a function of tube length $L$. The mean free paths are determined from the inverse slopes of the least-square linear fits (dashed, black lines in the inset) calculated using an automated numerical routine at each frequency. The shaded regions in the main figure correspond to the 92.5\% confidence interval for the slope. Below 0.25 THz, the confidence interval is very large (not shown) due to numerical uncertainties, inhibiting the reliable determination of mean free paths for very small frequencies.}  
\label{fig:mfp1}
 \end{center}
\end{figure}

The MFPs obtained by fitting $\Lambda(\omega)$ to Eq. \eqref{eq:T_diffusive} are shown in Fig. \ref{fig:mfp1}. The inset demonstrates the fitting procedure, where $M(\omega)/\ca{T}(\omega)$ has been calculated for tube lengths $L=0.5$ nm, $L=200$ nm, $L=400$ nm, $L=600$ nm, $L=800$ nm, $L=1$ $\mu$m, $L=2$ $\mu$m and $L=4$ $\mu$m. The values of $\Lambda(\omega)$ are obtained from the inverse slope of the linear fit to the data points and they are independent of the tube length. The shaded regions in Fig. \ref{fig:mfp1} reflect the 92.5\% confidence interval for the slope. Figure shows that the MFP at high frequencies $f>20$ THz is around $\Lambda\sim 10$--$100$ nm, reflecting the strong reduction of the transmission in this frequency range for a tube of length $L=200$ nm. At low frequencies, however, MFP is longer and exceeds one micrometer below $2.9$ THz, reaching $\Lambda(\omega)\approx 25$ $\mu$m for $f=0.25$ THz. Between 1 THz and 18 THz, the MFP can be seen to scale as $\Lambda(\omega)\sim \omega^{-0.97}$. 

\begin{figure}[tb]
 \begin{center}
   \includegraphics[width=8.6cm]{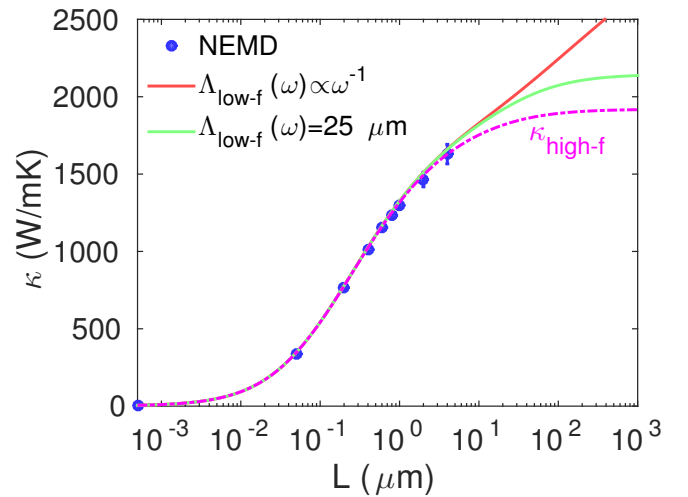}  
   \caption{(Color online) Thermal conductivity in semi-logarithmic scale. NEMD results (circles) are compared to the estimates $\kappa=\kappa_{\textrm{low-}f}+\kappa_{\textrm{high-}f}$ and $\kappa=\kappa_{\textrm{high-}f}$ for two different forms of $\kappa_{\textrm{low-}f}$. Here $\kappa_{\textrm{low-}f}$ and $\kappa_{\textrm{high-}f}$ are, respectively, the contributions of frequencies below and above $0.25$ THz to the integral of Eq. \eqref{eq:kappa_decomp}. The dash-dotted line includes only the high-frequency contribution calculated numerically by using the MFP data of Fig. \ref{fig:mfp1}. The low-frequency contribution has been calculated analytically assuming either that $\Lambda(\omega)\propto \omega^{-1}$ or $\Lambda(\omega)=25$ $\mu$m for low frequencies.}
\label{fig:kappa}
 \end{center}
\end{figure}

To see how the long mean free paths of Fig. \ref{fig:mfp1} affect the conductivity, Fig. \ref{fig:kappa} shows TC $ \kappa = QL/(A\Delta T)$
determined from NEMD simulations as a function of tube length $L$. The total average heat current $Q$ can be determined either by integrating the spectral heat current $q(\omega)$ over the positive frequencies or from the average power exchange with the heat baths. Figure shows that TC increases as a function of tube length $L$ even up to $L=4$ $\mu$m, the longest simulated tube length. The spectral decomposition \eqref{eq:kappa_decomp_q} of TC shows that the low-frequency phonons ($f\lesssim 3$ THz) are primarily responsible for the increase of conductivity in the longest simulated tubes (not shown). 

Because simulations for long tubes are very time-consuming, we have estimated TC for arbitrary $L$ by using the MFPs shown in Fig. \ref{fig:mfp1} and the spectral decomposition \eqref{eq:kappa_decomp} of TC. Figure \ref{fig:kappa} also illustrates how the low [$0$,$0.25$] THz and high [$0.25$,$\infty$) THz frequency parts of the integral contribute to the TC by separately showing the high-frequency contribution $\kappa_{\textrm{high-}f}$ and the total $\kappa=\kappa_{\textrm{low-}f}+\kappa_{\textrm{high-}f}$, where MFP in $\kappa_{\textrm{low-}f}$ has been assumed to scale as $\Lambda(\omega)\propto \omega^{-1}$ or $\Lambda(\omega)=25$ $\mu$m. The decomposition is discussed in more detail in App. \ref{sec:app1}. The conductivity $\kappa_{\textrm{high-}f}$ due to high-frequency phonons predicts the conductivity very well for all the simulated tube lengths, but the contribution of the low-frequency component $\kappa_{\textrm{low-}f}$ determines the scaling of TC for longer lengths. This contribution depends sensitively on the exact spectral form of low-frequency MFP, which we cannot extract reliably from the current simulations. If $\Lambda(\omega)$ scales as $\Lambda(\omega)\approx v/\omega$ for $\omega\to 0^+$, TC diverges logarithmically. If MFP diverges more slowly or tends to a constant, $\kappa$ converges as shown in Fig. \ref{fig:kappa}. We note that although TC of one-dimensional chains is known to diverge following a power-law \cite{mai07}, it is unknown if CNTs (or other physical systems) are truly one-dimensional in this respect. More experiments and simulations for longer tubes are still needed to settle the length-scaling of TC in long tubes.

\subsection{Wavenumber decomposition and relaxation times}
\label{sec:ktheta_results}
\begin{figure}[tb]
 \begin{center}
   \includegraphics[width=8.6cm]{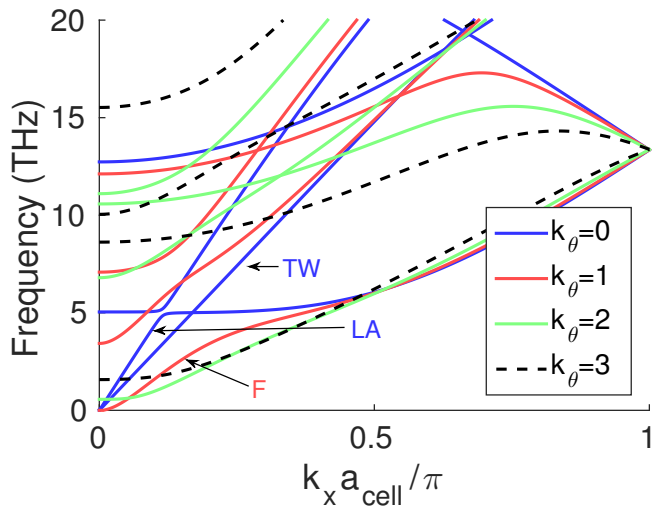}
  \caption{Wavenumber-decomposed low-frequency bandstructure of a (10,10) CNT calculated using the optimized Tersoff potential \cite{lindsay10}. At low frequencies, the longitudinal acoustic (LA) and twist (TW) modes with $k_{\theta}=0$ have linear dispersion. The flexural (F) mode with quadratic dispersion is doubly degenerate, with the two branches corresponding to $k_{\theta}=\pm 1$. The wavevector $k_x$ along the tube axis is scaled by $\pi/a_{\textrm{cell}}$, where $a_{\textrm{cell}}=\sqrt{3}a_{\textrm{CC}}$ and $a_{\textrm{CC}}$ is the nearest-neighbor carbon-carbon distance.}  
\label{fig:bs}
 \end{center}
\end{figure}

We now turn to the calculation of the transmission functions and mean free paths for different angular wavenumbers $k_{\theta}$. The phonon bandstructure of a (10,10) nanotube has been calculated earlier by Ong and Pop \cite{ong10} for different $k_{\theta}$ from the spectral energy density. For completeness, we show the decomposed bandstructure in Fig. \ref{fig:bs}, calculated here directly by determining the spatial Fourier transform of the wavenumber-decomposed force constant matrix (defined in App. \ref{app:ktheta}) along the tube axis and diagonalization. As shown earlier \cite{popov00,mahan04}, the longitudinal acoustic (LA) and twist (TW) modes, which have linear dispersion $\omega \sim k_x$ for small wavevector $k_x$, are constant in amplitude along the tube circumference and therefore belong to the $k_{\theta}=0$ branch. The flexural (F) modes, which obey the quadratic dispersion law $\omega \sim k_x^2$ at small wavevectors, can be seen to belong to the $k_{\theta}=\pm 1$ branches. By separately calculating the transmissions $\ca{T}(\omega,k_{\theta}=0)$ and $\ca{T}(\omega,k_{\theta}=1)$ from Eq. \eqref{eq:T_ktheta}, we can therefore separately study the damping of linear and quadratic modes.

\begin{figure}[tb]
 \begin{center}
  \includegraphics[width=8.6cm]{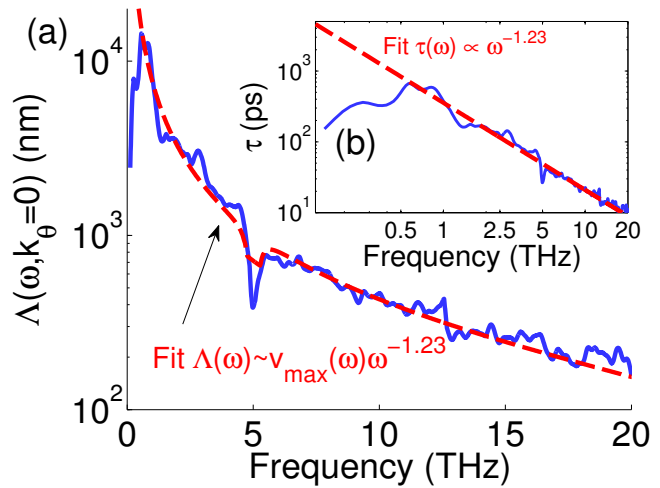}
  \caption{(Color online) (a) Mean free path $\Lambda(\omega,k_{\theta})$ for $k_{\theta}=0$. Below $5$ THz, only the acoustic LA and TW modes can be excited (see Fig. \ref{fig:bs}). The mean free path has been determined from Eq. \eqref{eq:mfp_ktheta} for $L=1$ $\mu$m. The relaxation time $\tau(\omega,k_{\theta})$ shown in (b) is well fitted by the power-law $\tau(\omega,k_{\theta}=0)\sim \omega^{-1.23}$ (dashed line).}  
\label{fig:100914f_Lambda_alpha0}
 \end{center}
\end{figure}

The MFPs $\Lambda(\omega,k_{\theta}=0)$ are shown in Fig. \ref{fig:100914f_Lambda_alpha0}(a). To save computational resources, we calculated the decomposed transmission functions $\ca{T}(\omega,k_{\theta})$ (shown for selected lengths in App. \ref{app:ktheta}) for a single length $L=1$ $\mu$m and determined MFP directly from Eq. \eqref{eq:mfp_ktheta} instead of performing calculations for multiple lengths and performing linear fitting as above. The simulation time duration was increased to $t_{\textrm{simu}}=100$ ns to enhance the statistical accuracy, allowing for reducing the width of the Gaussian smoothing window to $\Delta f=0.05$ THz for better resolution at low frequencies. Figure \ref{fig:100914f_Lambda_alpha0}(a) shows that the MFPs of LA and TW modes are strongly frequency-dependent at low frequencies and extend up to $10$ $\mu$m. 

To compare our results to the phonon life-times determined from EMD simulations, we also show the frequency-dependent relaxation time $\tau(\omega,k_{\theta})=\Lambda(\omega,k_{\theta})/v(\omega,k_{\theta})$ in Fig. \ref{fig:100914f_Lambda_alpha0}(b). Because there are typically multiple phonon branches (index by integer $p$) propagating at each $\omega$ and $k_{\theta}$, one needs to choose which group velocity $v_p(\omega,k_{\theta})$ to use in calculating $\tau(\omega,k_{\theta})$. We choose to use the maximum group velocity $v_{\textrm{max}}(\omega,k_{\theta})=\max_p v_p(\omega,k_{\theta})$ as the group velocity $v(\omega,k_{\theta})$ at each frequency $\omega$. We have checked that very similar results would be obtained by employing the average group velocity at each frequency $\omega$ in the calculation of $\tau(\omega,k_{\theta})$.

Figure \ref{fig:100914f_Lambda_alpha0}(b) shows that the relaxation time for $k_{\theta}=0$ decreases linearly in the log-log axis as a function of frequency for $f\gtrsim 0.6$ THz. Linear regression delivers the fit $\tau(\omega,k_{\theta}=0)\sim \omega^{-1.23}$ (dashed line). Below $f\lesssim 0.6$ THz, however, the power-law breaks down, suggesting that the divergence does not extend all the way to $\omega\to 0^+$. Because of the difficulties in determining the mean free paths at the very low frequencies, we cannot, however, certainly conclude if the break-down of the power-law is a physical phenomenon or a numerical artefact.

\begin{figure}[tb]
 \begin{center}
  \includegraphics[width=8.6cm]{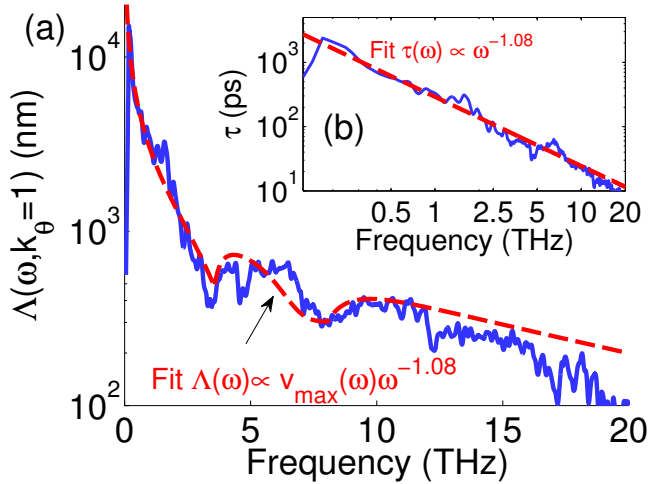}
  \caption{(Color online) (a) Mean free path $\Lambda(\omega,k_{\theta})$ for $k_{\theta}=1$, corresponding solely to the flexural (F) mode below $f\lesssim 3$ THz (see Fig. \ref{fig:bs}). The mean free path has been determined from Eq. \eqref{eq:mfp_ktheta} for $L=1$ $\mu$m. The relaxation time $\tau(\omega,k_{\theta}=1)$ shown in (b) is well fitted by the power-law $\tau(\omega,k_{\theta}=1)\sim \omega^{-1.08}$ (dashed line) above $0.2$ THz.}  
\label{fig:100914f_Lambda_alpha1}
 \end{center}
\end{figure}

Figure \ref{fig:100914f_Lambda_alpha1} shows the MFP $\Lambda(\omega,k_{\theta})$ and the relaxation time $\tau(\omega,k_{\theta})$ for the angular wavenumber $k_{\theta}=1$, corresponding to the F mode at low frequency. Below $10$ THz, the relaxation time $\tau(\omega,k_{\theta}=1)$ follows the power-law $\tau(\omega,k_{\theta}=1)\sim \omega^{-1.08}$ down to $f\approx 0.2$ THz. Below 0.2 THz, the power-law seems to break down, but because the low-frequency MFPs are prone to numerical uncertainties, we cannot rule out the possibility that the relaxation time of the flexural mode could actually diverge.

The found scaling laws for $\tau(\omega,k_{\theta})$ differ from the traditional Umklapp scattering life-time $\tau_U(\omega)\sim \omega^{-2}$, which has been applied in multiple works to estimate the length-dependence of TC in CNTs \cite{chantrenne04,wang06_apl,wang07_nanotech}, but which may not be applicable for CNTs due to the strict selection rules for phonon-phonon scattering  \cite{gu07,lindsay09}. EMD simulations \cite{ong11} have suggested $\tau(\omega) \sim \omega^{-1.1}$ scaling for the TW phonon life-time in a (10,10) tube, which is close to $\tau(\omega,k_{\theta}=0)\sim \omega^{-1.23}$ we found from NEMD. On the other hand, the same equilibrium simulations also predict $\tau(\omega) \sim \omega^{-2}$ for the flexural (F) mode, in contrast to $\tau(\omega,k_{\theta}=1) \sim \omega^{-1.08}$ found from NEMD. 

These differences demonstrate that the relaxation times determined from NEMD are not directly comparable to the relaxation times determined from EMD or first principles calculations: the latter methods reflect the total scattering rate in thermal equilibrium, neglecting the different roles of normal and Umklapp processes in generating thermal resistance and their complicated interplay in non-equilibrium situations. Considering that thermal transport is inherently a non-equilibrium process, we expect that the relaxation times determined from the NEMD simulations are the actually relevant scattering times that can be used to make predictions of, say, the length-dependence of TC. Quantum effects, which are expected to reduce the scattering rates particularly at low temperatures, could also be partially included in the NEMD method by replacing the classical heat baths employed in this work by quantum heat baths \cite{wang07}. 

\section{Conclusion} 
We have determined the frequency-dependent transmission function and phonon mean free paths in carbon nanotubes from nonequilibrium molecular dynamics simulations. The calculations relied on determining the spectral heat current for different tube lengths. Because our simulations exclude both boundary and impurity scattering, the MFPs reflect the scattering length in infinitely long, pristine tubes. Our results showed that the MFPs are approximately proportional to $\omega^{-0.97}$ over a wide range of frequencies and exceed $10$ $\mu$m for the low-frequency ($f<0.5$ THz) phonons. This leads to a thermal conductivity that increases as a function of tube length even in tubes as long as $4$ $\mu$m. The determined MFPs can be used to accurately predict the thermal conductivity of tubes shorter than 4 $\mu$m and they also provide insight into the conductivity of longer tubes. Relaxation times of selected phonon modes were shown to obey power-laws as a function of frequency, with generally different exponents than found using equilibrium simulations.

The presented methods for determining the contributions of different vibrational frequencies to thermal transfer are expected to be very useful in thermal engineering of carbon nanotube devices. Such calculations can be expected to improve, for example, the efficiency of thermoelectric materials by guiding the engineering process aiming at enhancing the contact to the heat source and sink. The method can also deliver transparent picture of the effect of non-linearities on thermal transfer, which is vital in enhancing the performance of non-linear thermal devices such as thermal diodes \cite{chang06}. More efficient design of such non-linear devices could eventually enable information processing using phonons \cite{li12_rmp}.

\section{Acknowledgements} 
We thank Shiyun Xiong for useful discussions. The computational resources were provided by the Finnish IT Center for Science and Aalto Science-IT project. The work was partially funded by the Aalto Energy Efficiency Research Programme (AEF) and the Academy of Finland.

\appendix
\section{Spectral decomposition and the length-dependence of thermal conductivity}
\label{sec:app1}
In this appendix, we discuss the separation of the thermal conductivity (TC) into low and high-frequency components in more detail. As stated in Sec. \ref{sec:results}, we separate $\kappa=QL/(A\Delta T)$ into its low- and high-frequency components as $\kappa=\kappa_{\textrm{low-}f}+\kappa_{\textrm{high-}f}$ by using the spectral decomposition \eqref{eq:kappa_decomp}:
\begin{equation}
 \kappa_{\textrm{low-}f} = \frac{k_BL}{A} \int_0^{\omega_c} \frac{d\omega}{2\pi} \frac{M(\omega)}{1+L/\Lambda(\omega)} \label{eq:kappa_lowf}
\end{equation}
and 
\begin{equation}
 \kappa_{\textrm{high-}f} = \frac{k_BL}{A} \int_{\omega_c}^{\omega_{\textrm{max}}} \frac{d\omega}{2\pi} \frac{M(\omega)}{1+L/\Lambda(\omega)} .
\end{equation}
Here $\omega_c=2\pi f_c$ is the cut-off frequency between the low and high frequencies and $\omega_{\textrm{max}}=2\pi \times 50$ THz is the maximum vibrational frequency. The long-$L$ limit of the high-frequency contribution $ \kappa_{\textrm{high-}f}$ is simple to evaluate, because $\Lambda(\omega)$ obtained from NEMD simulations is bounded from above in the integration range $\omega_c \leq \omega \leq \omega_{\textrm{max}}$. Therefore, for $L\gg \max_{\omega\in[\omega_c,\omega_{\textrm{max}}]} \Lambda $, the high-frequency contribution converges to the value
\begin{equation}
 \kappa_{\textrm{high-}f} = \frac{k_B}{A} \int_{\omega_c}^{\omega_{\textrm{max}}} \frac{d\omega}{2\pi} M(\omega)\Lambda(\omega).
\end{equation}

The long-$L$ limit of the low-frequency part can be similarly evaluated if MFP is bounded from above. If this is not the case and MFP diverges at low frequencies as $\Lambda(\omega)=b\omega^{-\eta}$ with $\eta>0$, one can calculate the asymptotic behaviour of the low-frequency part \eqref{eq:kappa_lowf} by choosing the cut-off small enough (e.g. $f_c=0.5$ THz) so that only the four acoustic modes can be excited for $f<f_c$ so that $M(\omega)=4$. Then Eq. \eqref{eq:kappa_lowf} becomes  (for $\eta\neq 1$)
\begin{widetext}
 \begin{alignat}{2}
 \kappa_{\textrm{low-}f} &= \frac{4k_BL}{A} \int_0^{\omega_c} \frac{d\omega}{2\pi} \frac{1}{1+L\omega^{\eta}/b} \\
  &= \frac{4k_BL \omega_c}{2\pi A}   {}_2 F_1 \left(1,\frac{1}{\eta},1+\frac{1}{\eta},-\omega_c^{\eta} \frac{L}{b} \right) \\
  &= \frac{4k_B}{2\pi A}  \Gamma\left(1+\frac{1}{\eta}\right) \left[ b^{1/\eta} \Gamma\left(1-\frac{1}{\eta}\right)L^{1-1/\eta} \right. + \left. b \omega_c^{1-\eta} \Gamma\left(\frac{1}{\eta}-1\right) \Gamma\left(\frac{1}{\eta}\right)^{-2} \right] + \mathcal{O} \left(\frac{1}{L} \right), \label{eq:kappa_lowf_power}
\end{alignat}
\end{widetext}
where ${}_2F_1$ is the hypergeometric function \cite{arfken}, written in the third line using its asymptotic expansion for large $L$ and the gamma function. For $\eta=1$, the integral \eqref{eq:kappa_lowf} is
\begin{equation}
 \kappa_{\textrm{low-}f} = \frac{4k_Bb}{A} \ln\left(1+\frac{L\omega_c}{b} \right). \label{eq:kappa_lowf_log}
\end{equation}
Equations \eqref{eq:kappa_lowf_power} and \eqref{eq:kappa_lowf_log} then directly lead to the conclusion that TC converges for $\eta < 1$, diverges as $\kappa(L)\propto L^{1-1/\eta}$ for $\eta>1$ and diverges logarithmically for $\eta=1$. Similar connection between the low-frequency scattering rates and the divergence of TC has been earlier proposed using mode-coupling theory \cite{lepri03} and by treating the finite length of the tube as a source of boundary scattering \cite{mingo05_nanolett}.

\section{Derivation of the wavenumber decomposition of the transmission function}

\label{app:ktheta}

\begin{figure}[tb]
 \begin{center}
  \includegraphics[width=8.6cm]{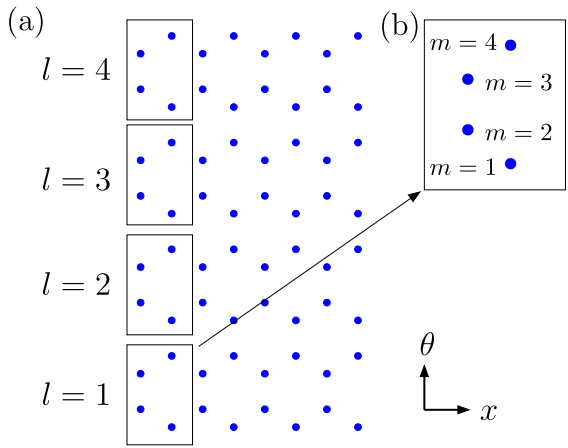}
  \caption{(a) Schematic description of the division of the $x$-translational unit cell in a (4,4) armchair CNT into four minimal unit cells (rectangles) labeled by the index $l$. The index $m\in \{1,2,3,4\}$ specifies the atom inside the minimal unit cell, as shown in (b). Note that we have plotted the CNT in two dimensions by choosing the vertical axis to correspond to be the azimuthal angle $\theta$ around the tube. The horizontal $x$-axis is the tube axis.}  
\label{fig:geom_theta}
 \end{center}
\end{figure}

We outline here the decomposition of the spectral heat current $q(\omega)$ into angular wavenumber components $q(\omega,k_{\theta})$ for an $(n,n)$ nanotube. We assume throughout that the velocities and the force constant matrix are represented in cylindrical coordinates to respect the rotational symmetry of the tube. The $x$-translational unit cell, which can be replicated along the tube axis to produce the whole CNT, contains $4n$ atoms. This unit cell can be divided into $n$ minimal translational unit cells, which repeat along the circumference as shown in Fig. \ref{fig:geom_theta}. Each atom in the $x$-translational unit cell can be labeled by its indices $l \in \{1,2,\dots,n\}$ and $m\in\{1,2,3,4\}$, which specify the minimal unit cell and the atom index inside the minimal cell, respectively. We denote the discrete Fourier transformed velocity vectors of the atoms belonging to the unit cells at the left and right sides of the imaginary plane separating the unit cells by $\hat{v}_{l,m,\alpha}^L(\omega)$ and $\hat{v}_{l,m,\alpha}^R(\omega)$, respectively, where the Greek index $\alpha$ stands for the co-ordinates $x$, $y$ and $z$. Each minimal unit cell $l$ is in angle $\theta_l=2\pi l/n+\theta_0$ with respect to the positive $y$-axis, where $\theta_0$ is a constant angle depending on the the chosen orientation of the positive $y$-axis. The transformation of the velocities to the angular wavenumber basis is then
\begin{equation}
 \tilde{v}_{k_{\theta},m,\alpha}^{L}(\omega) = \sum_{l \in \{1,\dots,n\}} e^{ik_{\theta} \theta_l} \hat{{v}}^{L}_{l,m,\alpha} (\omega),
\end{equation}
and similarly for $\hat{\bb{v}}_R$. The force constant matrix $\bb{K}$ specifying the force constants between the atoms located in these neighboring unit cells can be similarly transformed into the angular wavenumber basis as
\begin{equation}
 \tilde{K}^{k_{\theta}}_{m_1,\alpha;m_2,\beta} = \sum_{l_1 \in \{1,\dots,n\}} e^{ik_{\theta} (\theta_{l_1}-\theta_{l_2}) } {K}_{l_1,m_1,\alpha;l_2,m_2,\beta} \label{eq:K_ktheta}.
\end{equation}
where $l_2$ is arbitrary. With these decompositions at hand, it is straightforward to show that the decomposed spectral heat current is
\begin{widetext}
\begin{equation}
 \tilde{q}^{}(\omega,k_{\theta}) = -\frac{2}{\omega n t_{\textrm{simu}}} \sum_{m_1,m_2} \sum_{\alpha,\beta} \textrm{Im}\left\langle  {\tilde{v}}_{k_{\theta},m_1,\alpha}^L(\omega)^{*} \tilde{{K}}^{k_{\theta}}_{m_1,\alpha;m_2,\beta} {\tilde{v}}_{k_{\theta},m_2,\beta}^R(\omega) \right\rangle.
\end{equation}
\end{widetext}
\begin{figure}[t]
 \begin{center}
 \includegraphics[width=8.6cm]{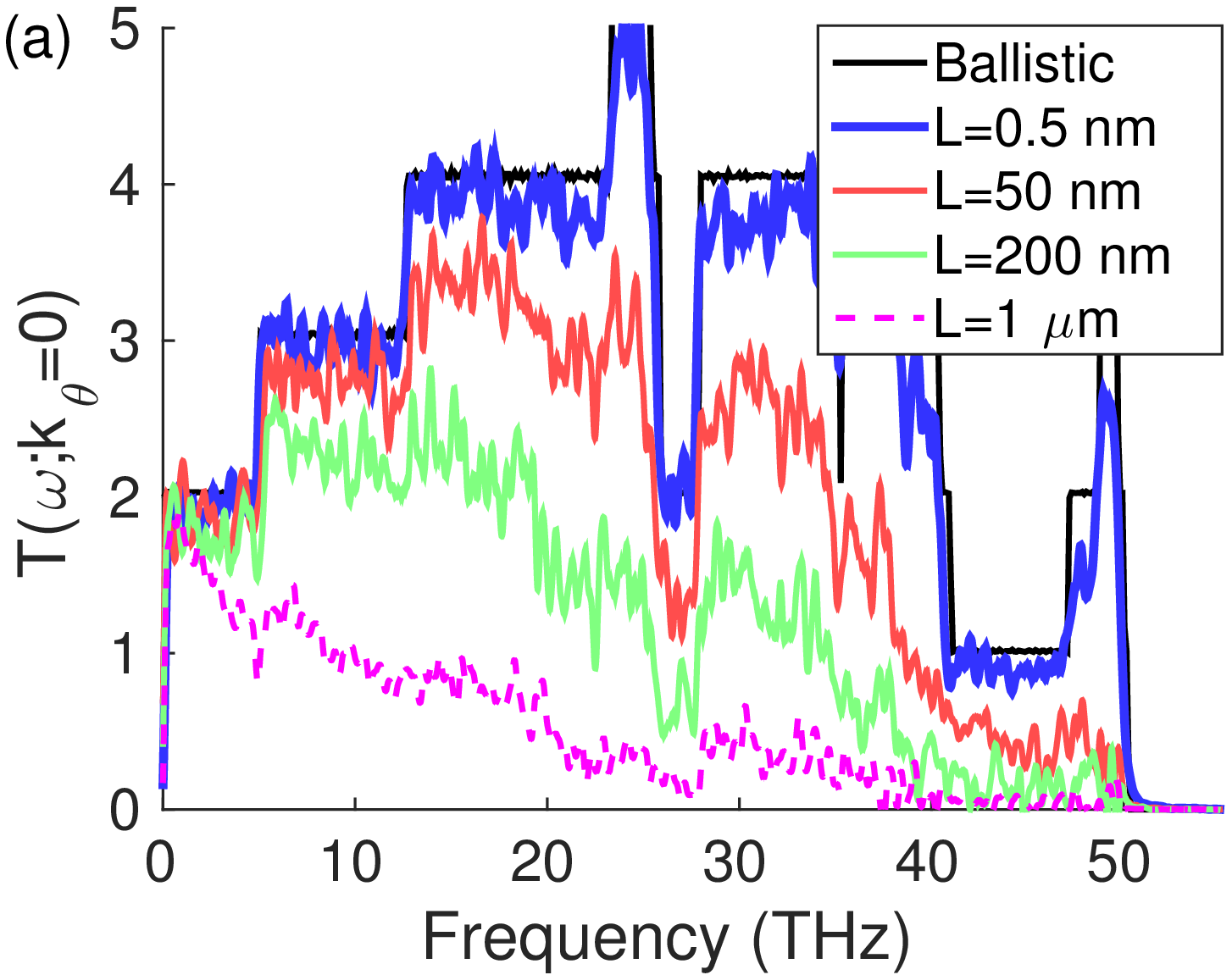}
 \includegraphics[width=8.6cm]{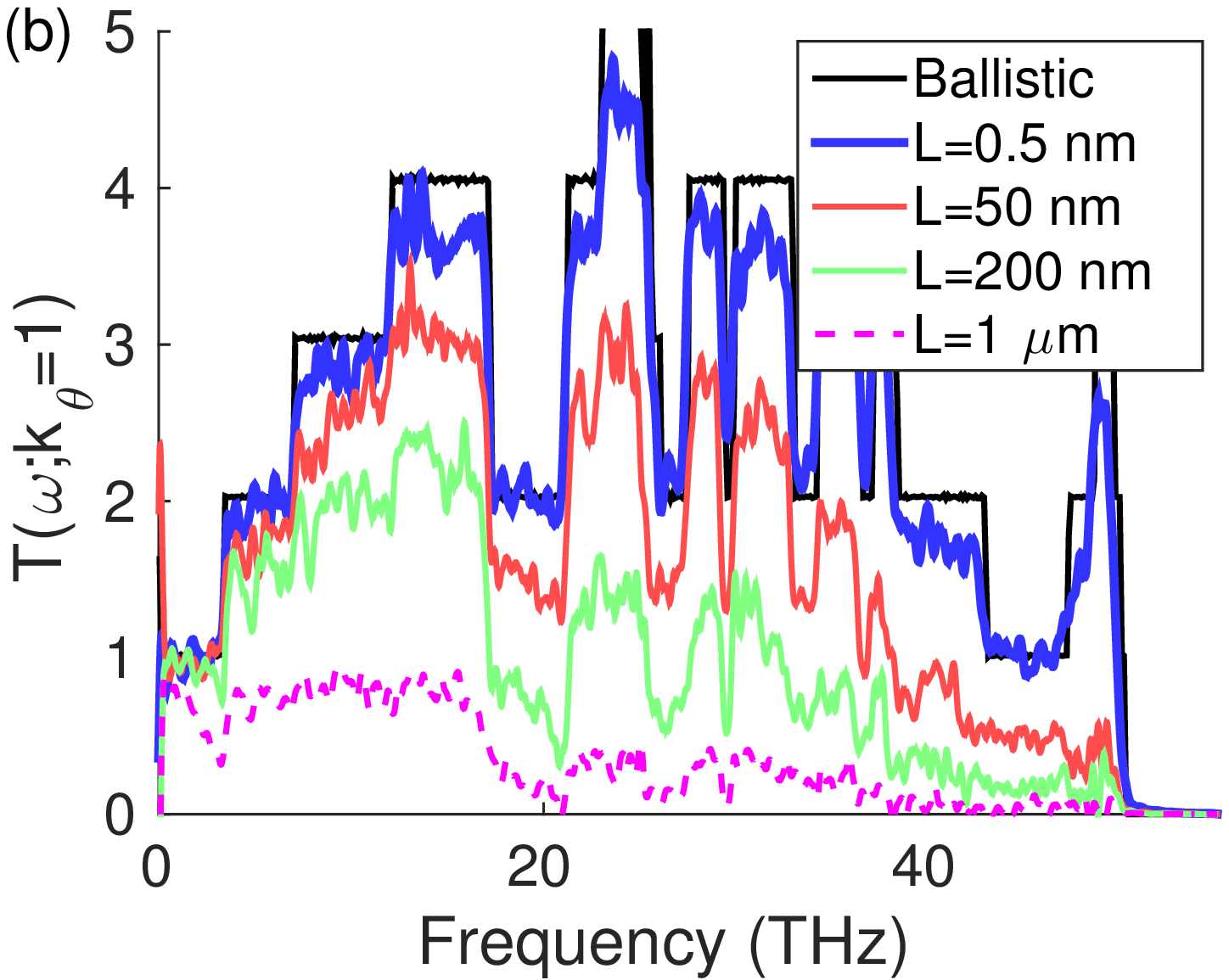}
  \caption{(Color online) Decomposed transmission function $\ca{T}(\omega;k_{\theta})$ for (a) $k_{\theta}=0$ and (b) $k_{\theta}=1$. At low frequencies, $k_{\theta}=0$ corresponds to the LA and TW modes, whereas $k_{\theta}=1$ corresponds to the flexural (F) mode.}  
\label{fig:Tom_alpha}
 \end{center}
\end{figure}

Figure \ref{fig:Tom_alpha} shows the wavenumber-decomposed transmission function \eqref{eq:T_ktheta} for (a) $k_{\theta}=0$ and (b) $k_{\theta}=1$. For $k_{\theta}=0$, only the LA and TW modes can be excited at low frequencies, so the ballistic transmission in Fig. \ref{fig:Tom_alpha}(a) is therefore equal to two at low frequencies. For $L=200$ nm, the low-frequency transmission is close to two due to the long mean free path of LA and TW modes. For $L=1$ $\mu$m, however, the transmission at $f\lesssim 5$ THz is already significantly smaller, suggesting that MFP is of the order of micrometer. The MFP can also be seen to decrease as a function of frequency.

For $k_{\theta}=1$ [Fig. \ref{fig:Tom_alpha}(b)], only the flexural F mode with quadratic dispersion $\omega \sim k_x^2$ can be excited below $f<3$ THz. The ballistic transmission is therefore equal to unity. Again, the transmission at low frequencies remains close to the ballistic value even for $L=200$ nm at $T=300$ K. For $L=1$ $\mu$m, however, the flexural mode is visibly dampened, with the damping increasing as a function of frequency.


\begin{thebibliography}{10}

\bibitem{berber00}
S.~Berber, Y.-K. Kwon, and D.~Tom\'anek,
\newblock Phys. Rev. Lett. {\bf 84}, 4613 (2000).

\bibitem{che00}
J.~Che, T.~Çagin, and W.~A.~G. III,
\newblock Nanotechnology {\bf 11}, 65 (2000).

\bibitem{osman01}
M.~A. Osman and D.~Srivastava,
\newblock Nanotechnology {\bf 12}, 21 (2001).

\bibitem{mingo05}
N.~Mingo and D.~A. Broido,
\newblock Phys. Rev. Lett. {\bf 95}, 096105 (2005).

\bibitem{donadio07}
D.~Donadio and G.~Galli,
\newblock Phys. Rev. Lett. {\bf 99}, 255502 (2007).

\bibitem{cao12}
A.~Cao and J.~Qu,
\newblock J. Appl. Phys. {\bf 112},  (2012).

\bibitem{salaway14}
R.~N. Salaway and L.~V. Zhigilei,
\newblock International Journal of Heat and Mass Transfer {\bf 70}, 954
  (2014).

\bibitem{kim01}
P.~Kim, L.~Shi, A.~Majumdar, and P.~L. McEuen,
\newblock Phys. Rev. Lett. {\bf 87}, 215502 (2001).

\bibitem{yu05}
C.~Yu, L.~Shi, Z.~Yao, D.~Li, and A.~Majumdar,
\newblock Nano Letters {\bf 5}, 1842 (2005).

\bibitem{pop05}
E.~Pop, D.~Mann, J.~Cao, Q.~Wang, K.~Goodson, and H.~Dai,
\newblock Phys. Rev. Lett. {\bf 95}, 155505 (2005).

\bibitem{pop06}
E.~Pop, D.~Mann, Q.~Wang, K.~Goodson, and H.~Dai,
\newblock Nano Letters {\bf 6}, 96 (2006).

\bibitem{marconnet13}
A.~M. Marconnet, M.~A. Panzer, and K.~E. Goodson,
\newblock Rev. Mod. Phys. {\bf 85}, 1295 (2013).

\bibitem{kaur14}
S.~Kaur, N.~Raravikar, B.~A. Helms, R.~Prasher, and D.~F. Ogletree,
\newblock Nature Comm. {\bf 5}, 3082 (2014).

\bibitem{prasher09}
R.~S. Prasher, X.~J. Hu, Y.~Chalopin, N.~Mingo, K.~Lofgreen, S.~Volz, F.~Cleri,
  and P.~Keblinski,
\newblock Phys. Rev. Lett. {\bf 102}, 105901 (2009).

\bibitem{gao10}
Y.~Gao, A.~Marconnet, M.~Panzer, S.~LeBlanc, S.~Dogbe, Y.~Ezzahri, A.~Shakouri,
  and K.~Goodson,
\newblock Journal of Electronic Materials {\bf 39}, 1456 (2010).

\bibitem{chang06_prl}
C.~W. Chang, D.~Okawa, H.~Garcia, A.~Majumdar, and A.~Zettl,
\newblock Phys. Rev. Lett. {\bf 99}, 045901 (2007).

\bibitem{chang06}
C.~W. Chang, D.~Okawa, A.~Majumdar, and A.~Zettl,
\newblock Science {\bf 314}, 1121 (2006).

\bibitem{balandin11}
A.~A. Balandin,
\newblock Nat. Mater. {\bf 10}, 569 (2011).

\bibitem{ju99}
Y.~S. Ju and K.~E. Goodson,
\newblock Appl. Phys. Lett. {\bf 74} (1999).

\bibitem{klemens94}
P.~Klemens and D.~Pedraza,
\newblock Carbon {\bf 32}, 735  (1994).

\bibitem{cao04}
J.~X. Cao, X.~H. Yan, Y.~Xiao, and J.~W. Ding,
\newblock Phys. Rev. B {\bf 69}, 073407 (2004).

\bibitem{mingo05_nanolett}
N.~Mingo and D.~A. Broido,
\newblock Nano Letters {\bf 5}, 1221 (2005),
\newblock PMID: 16178214.

\bibitem{gu07}
Y.~Gu and Y.~Chen,
\newblock Phys. Rev. B {\bf 76}, 134110 (2007).

\bibitem{lindsay09}
L.~Lindsay, D.~A. Broido, and N.~Mingo,
\newblock Phys. Rev. B {\bf 80}, 125407 (2009).

\bibitem{dames06}
C.~Dames and G.~Chen,
\newblock in {\em Thermoelectrics Handbook, Macro to Nano}, edited by D.~M.
  Rowe, Taylor \& Francis, New York, 2006.

\bibitem{dames13}
F.~Yang and C.~Dames,
\newblock Phys. Rev. B {\bf 87}, 035437 (2013).

\bibitem{minnich11a}
A.~J. Minnich, J.~A. Johnson, A.~J. Schmidt, K.~Esfarjani, M.~S. Dresselhaus,
  K.~A. Nelson, and G.~Chen,
\newblock Phys. Rev. Lett. {\bf 107}, 095901 (2011).

\bibitem{regner13}
K.~T. Regner, D.~P. Sellan, Z.~Su, C.~H. Amon, A.~J. McGaughey, and J.~A.
  Malen,
\newblock Nature Comm. {\bf 4}, 1640 (2013).

\bibitem{johnson13}
J.~A. Johnson, A.~A. Maznev, J.~Cuffe, J.~K. Eliason, A.~J. Minnich, T.~Kehoe,
  C.~M. Sotomayor-Torres, G.~Chen, and K.~A. Nelson,
\newblock Phys. Rev. Lett. {\bf 110}, 025901 (2013).

\bibitem{cuffe14}
J.~Cuffe, J.~K. Eliason, A.~A. Maznev, K.~C. Collins, J.~A. Johnson,
  A.~Shchepetov, M.~Prunnila, J.~Ahopelto, C.~M.~S. Torres, G.~Chen, and K.~A.
  Nelson,
\newblock arXiv:1408.6747, 2014.

\bibitem{ladd86}
A.~J.~C. Ladd, B.~Moran, and W.~G. Hoover,
\newblock Phys. Rev. B {\bf 34}, 5058 (1986).

\bibitem{mcgaughey04}
A.~J.~H. McGaughey and M.~Kaviany,
\newblock Phys. Rev. B {\bf 69}, 094303 (2004).

\bibitem{ong11}
Z.-Y. Ong, E.~Pop, and J.~Shiomi,
\newblock Phys. Rev. B {\bf 84}, 165418 (2011).

\bibitem{xiao03}
Y.~Xiao, X.~H. Yan, J.~X. Cao, and J.~W. Ding,
\newblock Journal of Physics: Condensed Matter {\bf 15}, L341 (2003).

\bibitem{hepplestone06}
S.~P. Hepplestone and G.~P. Srivastava,
\newblock Phys. Rev. B {\bf 74}, 165420 (2006).

\bibitem{ziman}
J.~Ziman,
\newblock {\em Electrons and Phonons: The Theory of Transport Phenomena in
  Solids} (Oxford University Press, USA, 2001).

\bibitem{Note1}
Recent publication has suggested that the textbook explanation of the role of
  normal and Umklapp processes is actually oversimplified: see A. A. Maznev and
  O.B. Wright, Am. J. Phys. \protect \textbf {82}, 1062 (2014).

\bibitem{saaskilahti14b}
K.~S\"a\"askilahti, J.~Oksanen, J.~Tulkki, and S.~Volz,
\newblock Phys. Rev. B {\bf 90}, 134312 (2014).

\bibitem{lepri03}
S.~Lepri, R.~Livi, and A.~Politi,
\newblock Phys. Rep. {\bf 377}, 1 (2003).

\bibitem{narayan09}
O.~Narayan and A.~P. Young,
\newblock Phys. Rev. E {\bf 80}, 011107 (2009).

\bibitem{hardy63}
R.~J. Hardy,
\newblock Phys. Rev. {\bf 132}, 168 (1963).

\bibitem{guajardo10}
A.~Guajardo-Cuéllar, D.~B. Go, and M.~Sen,
\newblock J. Chem. Phys. {\bf 132}, 104111 (2010).

\bibitem{rego98}
L.~G.~C. Rego and G.~Kirczenow,
\newblock Phys. Rev. Lett. {\bf 81}, 232 (1998).

\bibitem{datta}
S.~Datta,
\newblock {\em Electronic Transport in Mesoscopic Systems} (Cambridge
  University Press, 1997).

\bibitem{wang06_apl}
J.~Wang and J.-S. Wang,
\newblock Applied Physics Letters {\bf 88},  (2006).

\bibitem{yamamoto09}
T.~Yamamoto, S.~Konabe, J.~Shiomi, and S.~Maruyama,
\newblock Applied Physics Express {\bf 2}, 095003 (2009).

\bibitem{saaskilahti13}
K.~S\"a\"askilahti, J.~Oksanen, and J.~Tulkki,
\newblock Phys. Rev. E {\bf 88}, 012128 (2013).

\bibitem{savic08_prl}
I.~Savi\ifmmode~\acute{c}\else \'{c}\fi{}, N.~Mingo, and D.~A. Stewart,
\newblock Phys. Rev. Lett. {\bf 101}, 165502 (2008).

\bibitem{popov00}
V.~N. Popov, V.~E. Van~Doren, and M.~Balkanski,
\newblock Phys. Rev. B {\bf 61}, 3078 (2000).

\bibitem{mahan04}
G.~D. Mahan and G.~S. Jeon,
\newblock Phys. Rev. B {\bf 70}, 075405 (2004).

\bibitem{chalopin13}
Y.~Chalopin and S.~Volz,
\newblock Appl. Phys. Lett. {\bf 103}, 051602 (2013).

\bibitem{lindsay10}
L.~Lindsay and D.~A. Broido,
\newblock Phys. Rev. B {\bf 81}, 205441 (2010).

\bibitem{plimpton95}
S.~Plimpton,
\newblock J. Comput. Phys. {\bf 117}, 1  (1995).

\bibitem{lammps_website}
http://lammps.sandia.gov.

\bibitem{mai07}
T.~Mai, A.~Dhar, and O.~Narayan,
\newblock Phys. Rev. Lett. {\bf 98}, 184301 (2007).

\bibitem{ong10}
Z.-Y. Ong and E.~Pop,
\newblock Journal of Applied Physics {\bf 108},  (2010).

\bibitem{chantrenne04}
P.~Chantrenne and J.-L. Barrat,
\newblock Superlattices and Microstructures {\bf 35}, 173  (2004).

\bibitem{wang07_nanotech}
Z.~Wang, D.~Tang, X.~Zheng, W.~Zhang, and Y.~Zhu,
\newblock Nanotechnology {\bf 18}, 475714 (2007).

\bibitem{wang07}
J.-S. Wang,
\newblock Phys. Rev. Lett. {\bf 99}, 160601 (2007).

\bibitem{li12_rmp}
N.~Li, J.~Ren, L.~Wang, G.~Zhang, P.~H\"anggi, and B.~Li,
\newblock Rev. Mod. Phys. {\bf 84}, 1045 (2012).

\bibitem{arfken}
G.~Arfken,
\newblock {\em Mathematical Methods for Physicists}, 3rd ed. (Orlando, FL:
  Academic Press, 1985).

\end{thebibliography}
\end{document}